\documentstyle[12pt,epsf]{article}

\renewcommand{\cite}[1]{$^{\ref{#1}}$}

\newcommand{\beq}{\begin{equation}}
\newcommand{\eeq}{\end{equation}}
 \catcode`@=11
\def\@cite#1#2{\unskip\nobreak\relax
    \def\@tempa{$\m@th^{\hbox{\the\scriptfont0 #1}}$}%
    \futurelet\@tempc\@citexx}
\def\@citexx{\ifx.\@tempc\let\@tempd=\@citepunct\else
    \ifx,\@tempc\let\@tempd=\@citepunct\else
    \let\@tempd=\@tempa\fi\fi\@tempd}
\def\@citepunct{\@tempc\edef\@sf{\spacefactor=\the\spacefactor\relax}\@tempa
    \@sf\@gobble}

\def\citenum#1{{\def\@cite##1##2{##1}\cite{#1}}}
\def\citea#1{\@cite{#1}{}}

\newcount\@tempcntc
\def\@citex[#1]#2{\if@filesw\immediate\write\@auxout{\string\citation{#2}}\fi
  \@tempcnta\z@\@tempcntb\m@ne\def\@citea{}\@cite{\@for\@citeb:=#2\do
    {\@ifundefined
       {b@\@citeb}{\@citeo\@tempcntb\m@ne\@citea\def\@citea{,}{\bf ?}\@warning
       {Citation `\@citeb' on page \thepage \space undefined}}%
    {\setbox\z@\hbox{\global\@tempcntc0\csname b@\@citeb\endcsname\relax}%
     \ifnum\@tempcntc=\z@ \@citeo\@tempcntb\m@ne
       \@citea\def\@citea{,}\hbox{\csname b@\@citeb\endcsname}%
     \else
      \advance\@tempcntb\@ne
      \ifnum\@tempcntb=\@tempcntc
      \else\advance\@tempcntb\m@ne\@citeo
      \@tempcnta\@tempcntc\@tempcntb\@tempcntc\fi\fi}}\@citeo}{#1}}
\def\@citeo{\ifnum\@tempcnta>\@tempcntb\else\@citea\def\@citea{,}%
  \ifnum\@tempcnta=\@tempcntb\the\@tempcnta\else
   {\advance\@tempcnta\@ne\ifnum\@tempcnta=\@tempcntb \else \def\@citea{--}\fi
    \advance\@tempcnta\m@ne\the\@tempcnta\@citea\the\@tempcntb}\fi\fi}
\catcode`@=12

\def\thebibliography#1{\section*{{{\normalsize
\bf References }
\rule{0pt}{0pt}}\@mkboth
  {REFERENCES}{REFERENCES}}\list
  {{\arabic{enumi}.}}{\settowidth\labelwidth{{#1}}%
    \leftmargin\labelwidth  \frenchspacing
    \advance\leftmargin\labelsep
    \itemsep=-0.2cm
    \usecounter{enumi}}
    \def\newblock{\hskip .11em plus .33em minus -.07em}
    \sloppy
    \sfcode`\.=1000\relax}

\begin{document}
\baselineskip=14pt
\thispagestyle{empty}

\font\fortssbx=cmssbx10 scaled \magstep2
\hbox to \hsize{
%\special{psfile=/NextLibrary/TeX/tex/inputs/uwlogo.ps
%			      hscale=8000 vscale=8000
%			       hoffset=-12 voffset=-2}
\hskip.5in \raise.1in\hbox{\fortssbx University of Wisconsin - Madison}
\hfill$\vcenter{\hbox{\bf MAD/PH/838}
            \hbox{June 1994}}$ }

\vspace{.3cm}

\begin{center}\footnotesize
{\bf THE DETECTION OF COLD DARK MATTER WITH NEUTRINO TELESCOPES}\footnote{Talk
presented by F.~Halzen at {\it MRST-94: "What Next?  Exploring the Future of
High-Energy Physics"}, McGill University, Montreal, Canada, May 1994.}\\[.2cm]
{F.~HALZEN and J.~E.~JACOBSEN}\\[2pt]
{\it Dept.\ of Physics, University of Wisconsin, Madison WI 53706}
\end{center}
\vspace{.1cm}

\begin{abstract}
High energy neutrinos are produced by the annihilation of dark matter
particles in our galaxy. These are presently searched for with large area, deep
underground neutrino telescopes. Cold dark matter particles, trapped inside the
sun, are an abundant source of such neutrinos.

The realization that astronomy and particle physics have independently
developed compelling arguments for new physics at the weak scale may have
defined the most important experimental challenge of our time. The existence of
 cold dark matter particles, interacting weakly with ordinary matter, has been
convincingly inferred from a range of astronomical data. It may not be
accidental that new particles at the weak scale are also required to revamp
Standard Model phenomenology into a consistent theoretical framework where
radiative corrections are under control. The lightest stable supersymmetric
particle, or neutralino, is an example of such a dark matter candidate. Its
mass is bracketed by a minimum value of order a few tens
of GeV, determined from unsuccessful accelerator searches, and a maximum value
of order 1~TeV imposed by particle physics as well as cosmological
constraints.

Back-of-the-envelope calculations are sufficient to demonstrate
how neutrino telescopes are competitive with existing and future
particle colliders such as the LHC in the search for weakly interacting massive
cold dark matter particles. We will emphasize that a $1\,\rm km^2$ area is the
natural scale for a future
instrument capable of probing the full GeV--TeV mass range of cold dark matter
particle candidates by searching for high energy neutrinos produced by their
annihilation in the sun. We speculate on what such a detector may look like.
\end{abstract}

\vspace{.1cm}

\section{Introduction}

It is believed that most of our Universe is made of cold dark matter
particles. Big bang cosmology implies that these particles have interactions
of order the weak scale, i.e.\ they are WIMPs.\cite{Seckel} We briefly review
the argument which is sketched in Fig.~1. In the early Universe WIMPs are in
equilibrium with photons. When the Universe cools to temperatures well below
the mass $m_\chi$ of the WIMP their density is Boltzmann-suppressed as
$\exp(-m_\chi/T)$. Their density today would be exponentially small if it were
not for the expansion of the Universe. At some point, as a result of this
expansion, WIMPs drop out of equilibrium with other particles and a relic
abundance persists. The mechanism is analogous to nucleosynthesis where the
density of helium and other elements is determined by competition between the
rate of nuclear reactions and the expansion of the Universe.

\begin{figure}
\centering
\epsfxsize=4in\hspace{0in}\epsffile{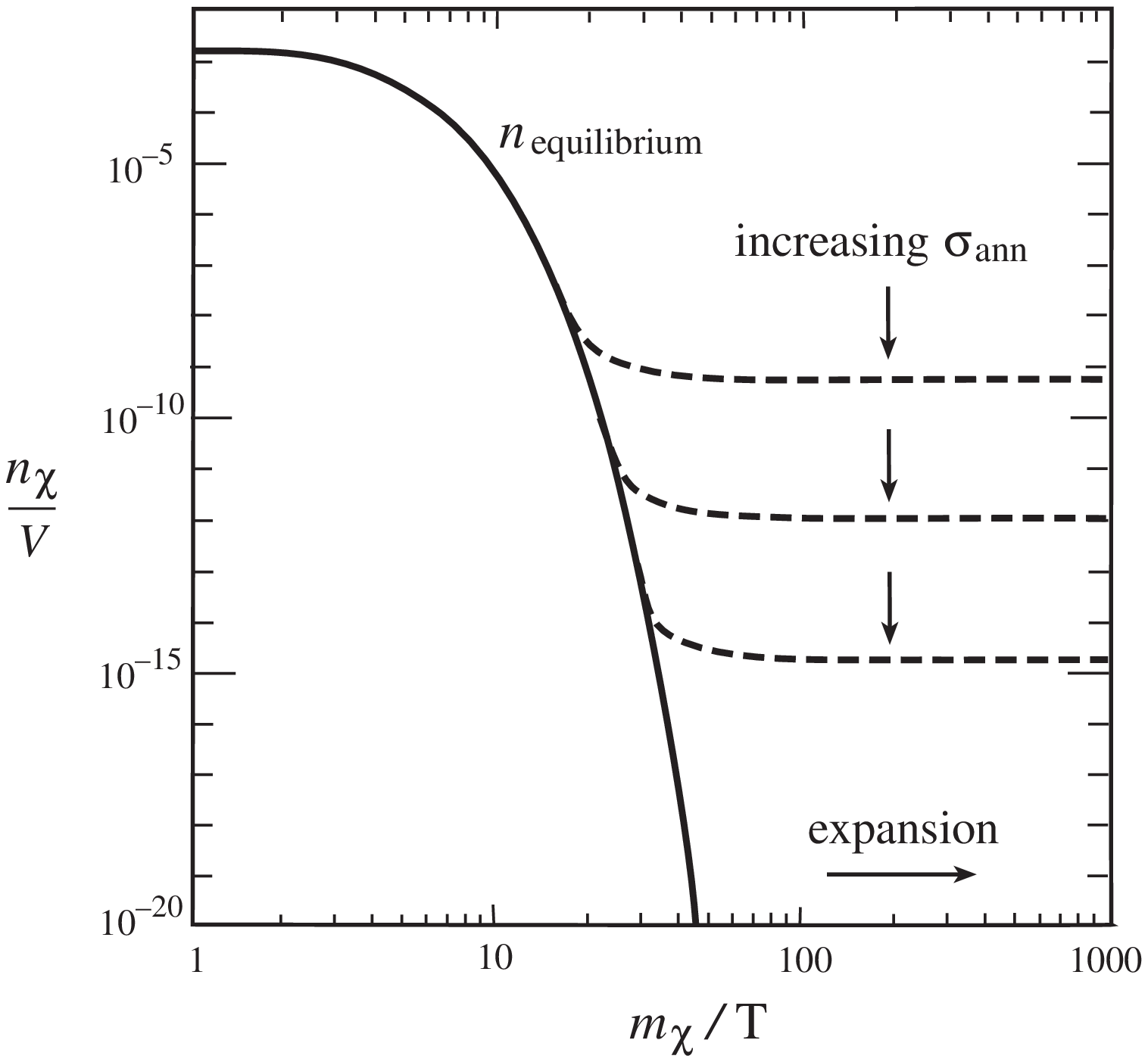}

\footnotesize Fig.~1
\end{figure}

At high temperatures WIMPS are abundant and they rapidly convert into lighter
particles. As long as they are in equilibrium lighter particles interact and
create WIMPs. The situation changes rapidly after the temperature drops below
the threshold for creating WIMPs, $T<m_\chi$, and their density falls
exponentially as a result of their annihilation into lighter particles. When
the expansion of the Universe has reduced their density to the point where
annihilation is no longer possible, a relic density ``freezes out'' which
determines the abundance of WIMPs today. This density is just determined by the
anihilation cross section; for a larger cross section freeze-out is delayed
resulting in a lower abundance today and vice versa. The scenario is sketched
in Fig.~1 where the density of WIMPs (in the comoving frame) is shown as a
function of time parametrized as the inverse of the temperature $T/m_\chi$.

For WIMPs to make up a large fraction of the Universe today, i.e.\ a large
fraction of $\Omega$, their annihilation cross section has to be ``just
right''. The annihilation cross section can be dimensionally written as $
\alpha^2/m_\chi^2$, where $\alpha$ is the fine-structure constant. It then
follows that
\begin{equation}
	\Omega\propto1/\sigma\propto m_\chi^2 \,. \label{omega}
\end{equation}
The critical point is that for $\Omega\simeq1$ we find that $m_\chi\simeq m_W$,
the mass of the weak intermediate boson. There is a deep connection between
critical cosmological density and the weak scale. Weakly interacting particles
remain to be discovered which constitute the bulk of the mass of the Universe.
It may not be an accident that the unruly behavior of radiative corrections in
the Standard Model also requires the existence of such particles, more about
that later.

When our galaxy was formed the cold dark matter inevitably clustered with the
luminous matter
to form a sizeable fraction of the
\begin{equation}
        \rho_{\chi}=0.4\rm~GeV/cm^3  \label{density}
\end{equation}
galactic matter density implied by observed rotation curves. Unlike the
baryons, the dissipationless WIMPs fill the galactic halo which is believed to
be an isothermal sphere of WIMPs with average velocity
\begin{equation}
         v_{\chi}=300\rm\ km/sec \,. \label{velocity}
\end{equation}

For a first look at the experimental problem of how to detect these particles
it is sufficient to recall that they are weakly interacting with masses in the
range
\begin{equation}
                \mbox{tens of GeV} < m_{\chi} < \rm TeV \,. \label{GT}
\end{equation}
Neutrinos produced by the annihilation of WIMPs represent the
experimental signature for the presence of halo dark matter. For the initial
discussion it is sufficient to assume that WIMPs, if sufficiently massive, will
predominantly annihilate into weak bosons
\begin{equation}
\chi + \bar \chi \to W^+ + W^- \,.
\end{equation}

It is amusing that we know almost everything about WIMPs; defining the
challenge of detecting them should be straightforward. Although other methods
for detecting WIMPs exist\cite{Seckel}, we will here discuss detectors which
exploit the indirect, but very striking neutrino signature of halo WIMP
annihilation. Heavy WIMPs annihilate in energetic neutrinos. These are easy to
detect because of their relatively large interaction cross section and the
large range of the muons they produce. Also, nature has been kind to us and
given us the sun as a nearby reservoir of WIMPs where their density is greatly
enhanced. A new generation of high energy neutrino telescopes are being
commissioned to meet the challenge.

\section{Neutrino Signature of WIMP-Annihilation in the Sun}

Galactic WIMPs, scattering off protons in the sun, loose energy. They may fall
below
escape velocity and be gravitationally trapped. Trapped dark matter particles
eventually come to equilibrium temperature, and therefore to rest at the
center of the sun. While the neutralino density builds up, their
annihilation rate into lighter particles increases until equilibrium is
achieved where the
annihilation rate equals half of the capture rate. The sun has thus become a
reservoir of WIMPs which annihilate into any open fermion, gauge boson
or Higgs channels. The leptonic decays from annihilation channels such as
$b\bar b$ heavy quark pairs and $W^+W^-$ turn the sun into a source of high
energy neutrinos.
Their energies are in the GeV
to TeV range, rather than in the familiar KeV to MeV range
from its nuclear burning.
These neutrinos can
be detected in deep underground experiments. Figure~2 shows a cartoon of the
chain of events leading from dark matter particles in the halo to a muon track,
pointing
back to the sun, in the Earth-based Cherenkov detector.

We will illustrate the power of neutrino telescopes as dark matter detectors
using as an example the search for a 500~GeV WIMP with a mass outside the
reach of present accelerator and future LHC experiments.
No long and complex code is necessary to qualitatively evaluate the potential
of high energy neutrino telescopes as dark matter detectors. For
quantitative calculations such code exists
and can be distributed to anyone interested.\cite{Halzen}

\begin{figure}
\centering
\epsfxsize=3.25in\hspace{0in}\epsffile{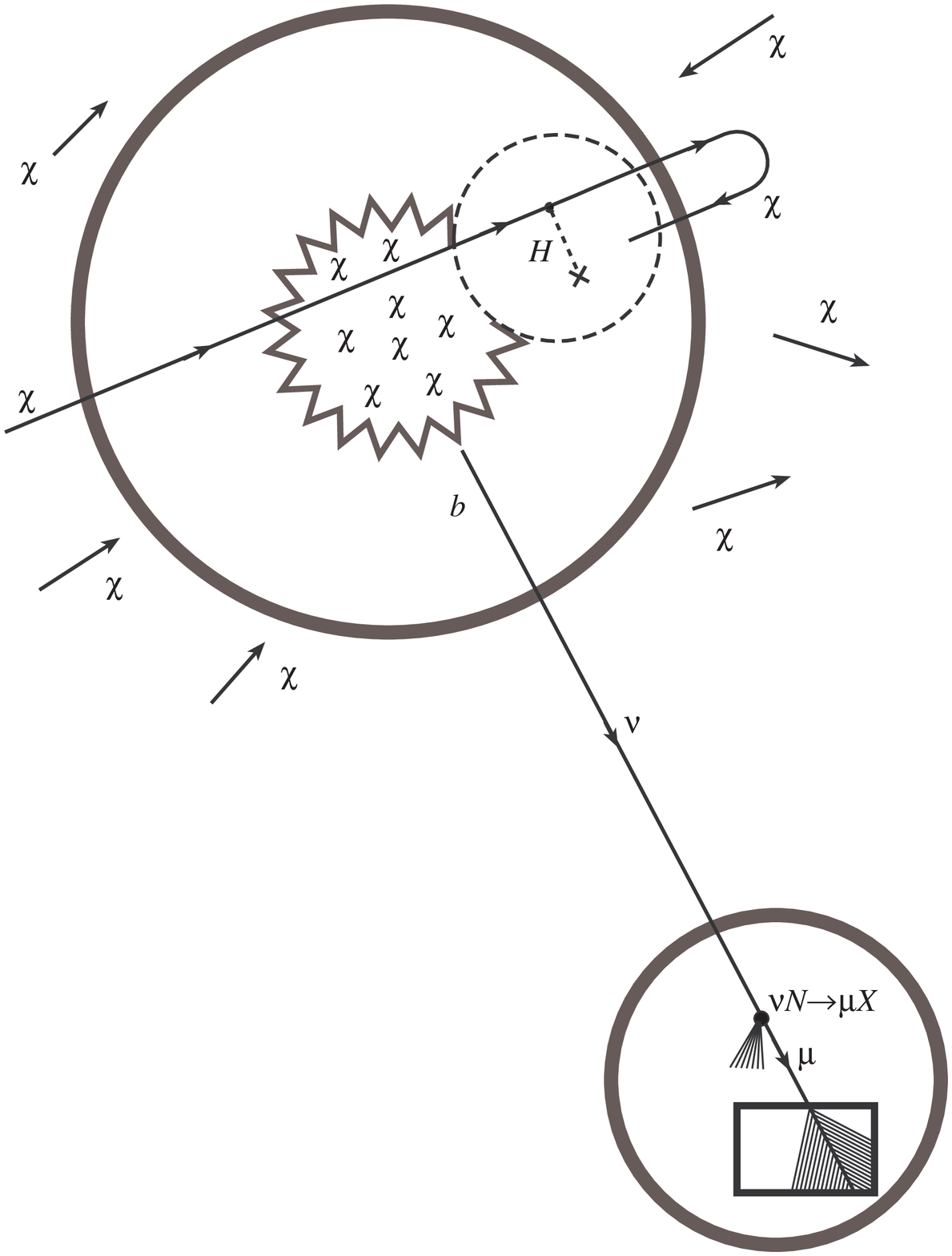}

\footnotesize Fig.~2
\end{figure}

A calculation of the rate of high energy muons of WIMP origin triggering
a detector is calculated in 5 easy steps.

\smallskip
\noindent
{\bf {\it Step 1:} The halo neutralino flux \boldmath{$\phi_{\chi}$} }

It is given by their number density (\ref{density}) and average velocity
(\ref{velocity}). From (\ref{density})
\begin{equation}
n_\chi = 8\times 10^{-4} \left[ 500{\rm\ GeV}\over m_\chi \right]\rm\
cm^{-3}
\end{equation}
and therefore
\begin{equation}
\phi_\chi = n_\chi v_\chi = 2\times 10^{4} \left[ 500{\rm\ GeV}\over m_\chi
\right] \rm\ cm^{-2}\, s^{-1} \,. \label{flux}
\end{equation}

\smallskip
\noindent
{\bf {\it Step 2:} Cross section \boldmath{$\sigma_{\rm sun}$} for the capture
of neutralinos by the
sun }

The probability that a WIMP is captured is proportional to the number of
target hydrogen nuclei in the sun (i.e.\ the solar mass divided by the nucleon
mass) and the WIMP-nucleon scattering cross section $\sigma(\chi N)$. We can
estimate the neutralino-nucleon cross section by
dimensional analysis. This yields $\left(G_F m_N^2\right)^2/m_Z^2$ which we can
envisage as the exchange of a neutral weak boson between the WIMP and a quark
in the nucleon. The main point is that the WIMP is known to be weakly
interacting. We obtain
for the solar capture cross section
\begin{equation}
\Sigma_{\rm sun} = n\sigma = {M_{\rm sun}\over m_N} \sigma(\chi N)
= \left[1.2\times 10^{57}\right] \left[10^{-41}\,\rm cm^2\right] \,.
\label{capture}
\end{equation}

\smallskip
\noindent
{\bf {\it Step 3:} Capture rate \boldmath{$N_{\rm cap}$} of neutralinos by the
sun }

$N_{\rm cap}$ is determined by the WIMP flux (\ref{flux}) and the
sun's capture
cross section (\ref{capture}) obtained in the first 2 steps:
\begin{equation}
N_{\rm cap} = \phi_\chi \Sigma_{\rm sun} = 3\times 10^{20}\,\rm s^{-1} \,.
\end{equation}

\smallskip
\noindent
{\bf {\it Step 4:} Number of solar neutrinos of dark matter origin }

One can check that the sun comes to a steady state where capture and
annihilation of WIMPs are in equilibrium. For a 500~GeV WIMP the
dominant annihilation rate is into weak bosons; each produces muon-neutrinos
with a leptonic branching ratio which is roughly 10\%:
\begin{equation}
\chi\bar\chi \to WW \to \mu \nu_\mu \,. \label{branch}
\end{equation}
Therefore, as we get 2 $W$'s for each capture, the number of neutrinos
generated in the sun is
\begin{equation}
N_\nu = {1\over 5} N_{\rm cap}
\end{equation}
and the corresponding neutrino flux at Earth is given by
\begin{equation}
 \phi_\nu = {N_\nu\over 4\pi d^2} = 2\times 10^{-8}\,\rm cm^{-2} s^{-1} \,,
\label{nu flux}
\end{equation}
where the distance $d$ is 1 astronomical unit.

\smallskip\goodbreak
\noindent
{\bf {\it Step 5:} Event rate in a high energy neutrino telescope }

It is evident that this flux is small enough to require the use of large
volumes of natural water or ice as a $\nu_\mu$ detection volume. The
secondary muons produced in the interaction provide the experimental signature
of the neutrinos in the underground detector; see Fig.~2. Optical modules view
the water or ice target and detect the muon by the Cherenkov light emitted as
it traverses the detector. By mapping the Cherenkov cone the solar origin of
the neutrino can be established, at least for the higher energies where muon
and neutrino directions are approximately aligned. The number of muons
observed in a detector of a given area is
\begin{equation}
N = \hbox{area}\int dE\,{dN_\nu\over dE}\,P_{\nu\to\mu} \,.
\end{equation}
The quantity $P_{\nu\to\mu}$ represents the combined probabilities that a
``solar'' neutrino interacts in the volume viewed by the optical modules and
that the muon, produced in the interaction, has a sufficient range to reach
the detector. It depends on the particle density of the target, the neutrino
interaction cross section and the range of the muon, therefore
\begin{equation}
P_{\nu\to\mu} = \rho_{\rm H_2O}\,\sigma_{\nu\to\mu}(E)\,R_\mu \,.
\label{P}
\end{equation}
Evaluating Eq.~(\ref{P}) only involves standard particle physics.

For (\ref{branch}) the $W$-energy is approximately $m_{\chi}$ and the
neutrino energy half that by 2-body kinematics. The energy of the detected
muon is given by
\begin{equation}
 E_\mu \simeq {1\over2} E_\nu \simeq {1\over4}m_\chi \,.
\end{equation}
where we used the fact that, in this energy range, roughly half of the neutrino
energy is transferred to the muon. Simple estimates of the neutrino
interaction cross section and the muon range can be obtained as follows
\begin{eqnarray}
  \sigma_{\nu\to\mu} &=& 10^{-38}\,{\rm cm^2}\,{E_\nu\over \rm GeV} =
2.5\times10^{-36}\,\rm cm^2  \label{sigma numu} \\
\noalign{\hbox{and}}
R_\mu &=& 5\ {\rm m}\,{E_\mu\over\rm GeV} = 625\ \rm m \,, \label{R mu}
\end{eqnarray}
which is the distance covered by a muon given that it loses 2~MeV for each
gram of matter traversed. We have now collected all the information to compute
the number of events in a detector of area $10^4\,$m$^2$, typical for those
presently under construction. For the neutrino flux given by (\ref{nu flux})
we obtain
\begin{equation}
 {\rm \#\ events/year = area} \times \phi_\nu \times \rho_{\rm H_2O} \times
\sigma_{\nu\to\mu} \times R_\mu \simeq 10 \,.
\end{equation}
Ten 125~GeV muons from the direction of the sun! It is a pretty safe bet that
such a signal will not be drowned by whatever real world experimental problems
dilute this naive estimate.

The above exercise is just meant to illustrate that present high energy
neutrino
telescopes compete with present and future accelerator experiments in the
search for dark matter and supersymmetry; see below. Especially for heavier
WIMPs the technique is very powerful because underground high energy neutrino
detectors have been
optimized to be sensitive in the energy region where the neutrino interaction
cross section and the range of the muon are large; notice the
$E$-factors in (\ref{sigma numu}), (\ref{R mu}). Also, for high energy
neutrinos the muon and neutrino are nicely aligned along a direction pointing
back to the sun with good angular resolution. Direct searches will have to
deliver detectors reaching better than 0.05~events/kg\,day sensitivity to
compete.\cite{Bottino}

In many places we made assumptions that lead to an underestimate of the
signal. We neglected, for instance, other decay channels contributing neutrinos
in (16). We did not
include the signal from annihilation of WIMPs trapped in the center of
the Earth.\cite{Gould}

\section{Now for those who like supersymmetry}

Particle physics provides us with rather compelling candidates for WIMPs. The
Standard Model is not a model, its radiative corrections are not under control.
A most elegant and economical way to revamp it
into a consistent and calculable framework is to make the model
supersymmetric. If supersymmetry is indeed Nature's extension of the Standard
Model it must produce new phenomena at or below the TeV scale. A very
attractive feature of supersymmetry is that it provides cosmology with a
natural dark matter candidate in form of a stable lightest supersymmetric
particle.\cite{Seckel} There are a priori six candidates: the (s)neutrino,
axi(o)n(o), gravitino and neutralino. These are, in fact, the only candidates
because supersymmetry completes the Standard Model all the way to the GUT
scale where its forces apparently unify. Because supersymmetry
logically completes the Standard Model with no other new physics threshold up
to the GUT-scale, it must supply the dark matter. So, if supersymmetry, dark
matter and accelerator detectors are on a level playing field. Here we will
focus on the neutralino which, along with the axion, is for various reasons
the most palatable WIMP candidate.\cite{Berezinsky}

The supersymmetric partners of
the photon, neutral weak boson and the two Higgs particles form four neutral
states, the lightest of which is the stable neutralino
\begin{equation}
\chi = z_{11} \tilde W_3 + z_{12} \tilde B + z_{13} \tilde H_1 + z_{14} \tilde
H_2 \,.
\end{equation}

In the minimal supersymmetric model (MSSM)\cite{Haber} down- and up-quarks
acquire mass by coupling to different Higgs particles, usually denoted by
$H_1$ and $H_2$, the lightest of which is required to have a mass of order the
$Z$-mass. Although the MSSM provides us with a definite calculational
framework, its parameters are many. For the present discussion we only have to
focus on the following terms in the MSSM lagrangian
\begin{equation}
L = \cdots \mu\tilde H_1 \tilde H_2 -{1\over2} M_1 \tilde B \tilde B
-{1\over2}M_2 \tilde W_3 \tilde W_3 - {1\over\sqrt2} g v_1 \tilde H_1 \tilde
W_3 - {1\over\sqrt2} g v_2 \tilde H_2 \tilde W_3 + \cdots \,,
\end{equation}
which introduce the (unphysical) masses $M_1,\ M_2$ and $\mu$ associated with
the neutral gauge bosons and Higgs particles, respectively. $M_1$ and $M_2$ are
related by the Weinberg angle. The lagrangian introduces two Higgs vacuum
expectation values $v_{1,2}$; the coupling $g$ is the known Standard Model
SU(2) coupling. Although the parameter space of the MSSM is more complex, a
first discussion of dark matter uses just 3 parameters:
\begin{equation}
                \mu,\ M_2,\ {\rm and}\ \tan\beta=v_2/v_1 \,.
\end{equation}
Further parameters which can also be varied include the masses of top, Higgs,
squarks, etc.

Neutralino masses less than a few tens of GeV have been excluded by
unsuccessful collider searches. For supersymmetry to resolve the
fine-tuning problems of the Standard Model the masses of supersymmetric
particles must be of order the weak scale and therefore, in practice, at the
TeV scale or below. Also, if neutralinos have masses of order a few TeV and
above, they overclose the Universe. Despite its rich parameter space
supersymmetry has therefore been framed inside a well defined GeV--TeV mass
window; see (\ref{GT}).

Assuming supersymmetry we can fill in ``the factors''  in the
``back-of-the-envelope'' estimates in the previous section. In supersymmetry,
heavy WIMPs annihilate indeed preferentially into weak bosons. Other important
annihilation
channels include \cite{Drees}
\begin{equation}
\chi + \bar\chi \to b + \bar b. \label{chi to b}
\end{equation}
Heavy quark decays dominate neutralino annihilation below the $WW$-threshold.

Also the dimensional estimate of  the neutralino-nucleon interaction cross
section $\sigma(\chi N)$ can be replaced by an explicit calculation. It
supports the dimensional estimate in the previous section. $\sigma(\chi N)$
receives contributions from 2 classes of diagrams: the exchange of Higgses and
weak bosons, and the exchange of squarks; see Fig.~3. The result is often
dominated by the large coherent cross section associated with the exchange of
the lightest Higgs particle $H_2$ and is of the form
\begin{equation}
\sigma = \alpha_H (G_F m_N^2)^2 {m_\chi^2\over(m_N + m_\chi)^2} \, {m_Z^2\over
m_H^4}
\end{equation}
or, for large $m_{\chi}$
\begin{equation}
\sigma = \alpha_H \left(G_F m_N^2\right)^2 {m_Z^2\over m_H^4} \,. \label{large}
\end{equation}
The proportionality parameter $\alpha_H$ is of order unity, but can become
as small as $10^{-2}$ in some regions of the MSSM parameter space. This is
illustrated in Fig.~4 where the MSSM parameter space is parametrized in terms
of the unphysical masses $M$($\mu$) of the unmixed wino(Higgsino). (The ratio
of the vacuum expectation values associated with the two Higgs particles
$v_2/v_1(=2)$ is here fixed to some arbitrary value.) The relation of these
parameters to the neutralino mass is shown in the figure. The full lines show
fixed values of the neutralino mass $m_{\chi}$. The lines labelled by squares
trace fixed values of the ``coupling'' $\alpha_H$. The dashed area
indicates $M$, $\mu$ values which are excluded by cosmological considerations.
In standard big bang cosmology neutralinos with the corresponding parameters
will overclose the Universe. Note that for a given $\chi$ mass there are two
possible states with the same $\alpha_H$ value. One of them will
preferentially annihilate into weak bosons, the other into fermions.
Therefore, their neutrino signature is provided by $W,Z$ decay and
semi-leptonic heavy
quark decays, respectively. Figure~4 illustrates that for heavy neutralinos,
which can only be searched for by the indirect methods discussed here and are
therefore of prime interest, any detector which can study dark matter with
$\alpha_H$ as small as 0.1 can exclude the bulk of the phase space currently
available to MSSM dark matter candidates.

\begin{figure}[t]
\centering
\epsfxsize=2.25in\hspace{0in}\epsffile{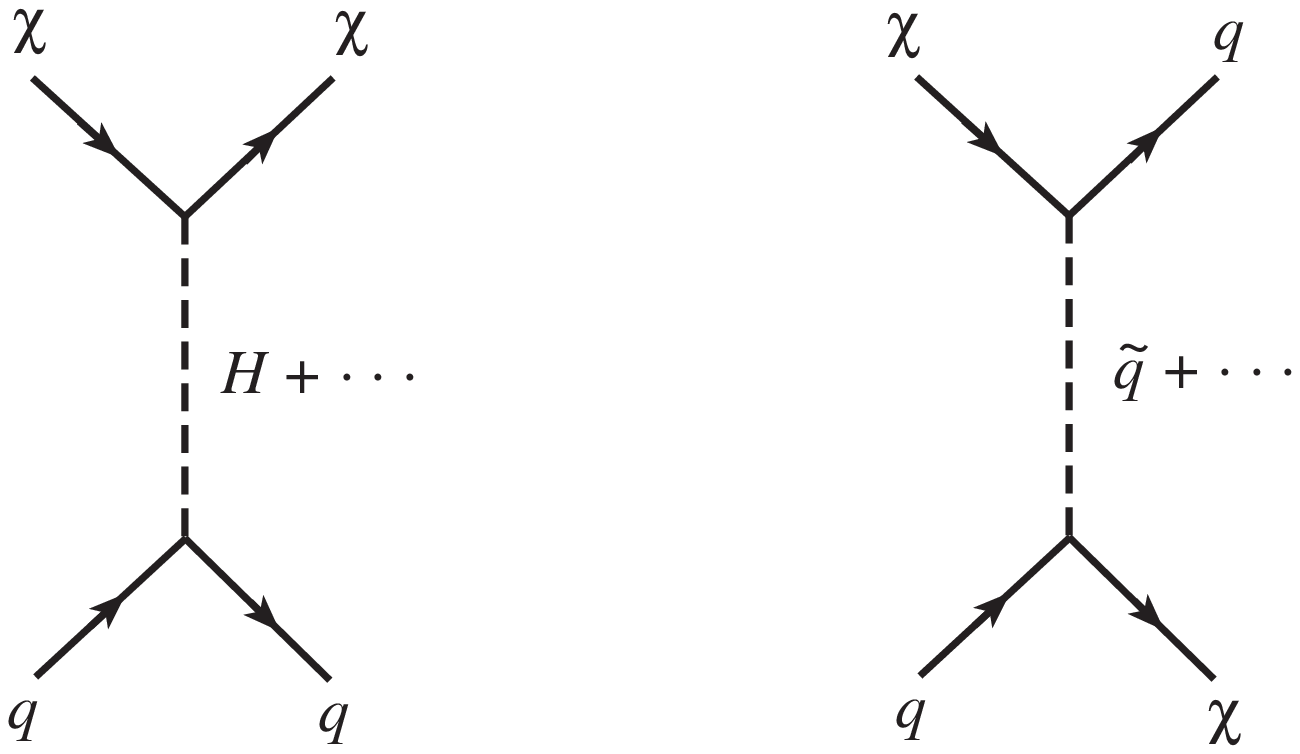}

\footnotesize Fig.~3. Neutralino-quark scattering by Higgs and squark exchange.

\vglue.5in
\epsfxsize=3.25in\hspace{0in}\epsffile{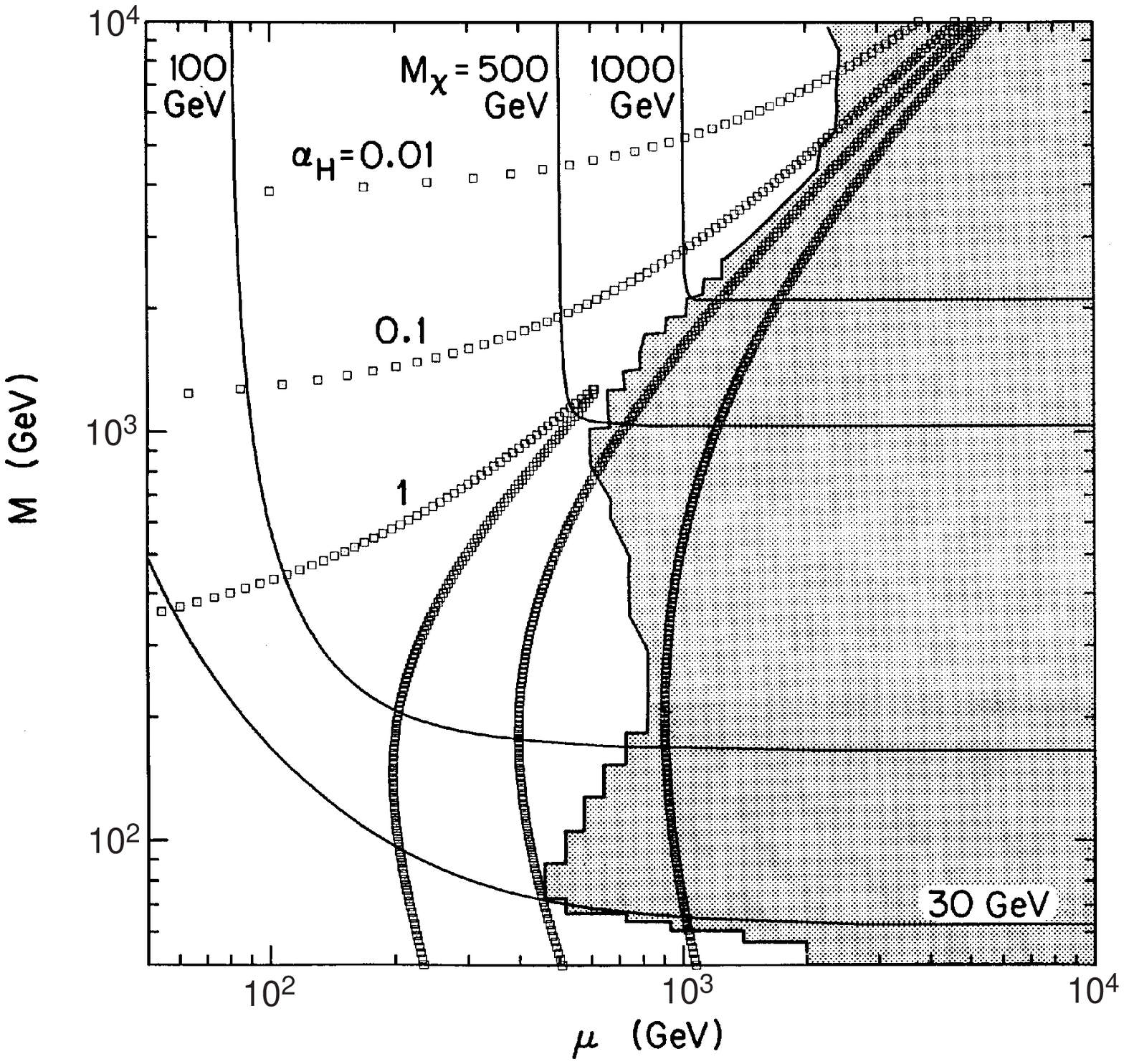}

\smallskip
\parbox{5.5in}{\footnotesize Fig.~4. Contours in the $M,\mu$ plane of constant
$\alpha_{H_2}=1.0,\
0.1,\ 0.01$ (boxes) and constant neutralino mass $M_\chi=30, 100, 500$ and
1000~GeV (solid). The shaded region is excluded by cosmological
considerations.}
\end{figure}

 Our main conclusions are summarized in Fig.~5 which exhibits, as a function of
the
neutralino mass, the detector area required to observe one event per year. The
detailed calculation confirms our previous estimate of 10 events per year for a
500 GeV neutralino.
The two branches in this and the following figures correspond to the two
solutions for a fixed neutralino mass; see also Fig.~4. Various annihilation
thresholds are clearly visible, most noticeable is the threshold associated
with the $W,Z$ mass near 100~GeV.  The graphs confirm that, realistically, a
detector of km-scale is required to study the full neutralino mass range. It
is clear from Fig.~5, however, that even detectors of more modest size can
radically improve on accelerator results.\cite{Mori}$^,$\cite{LoSecco}
Neutralinos
of 1~TeV mass are observable in a detector of area a few times
$10^3\,\rm m^2$. The
energy of the produced neutrinos is typically ``a fraction'' of the neutralino
mass, e.g.\ 1/2 for neutralino annihilation into a $W$ followed by a leptonic
$e \nu$ decay. For lower masses the event rates are small because the detection
efficiency for low energy neutrinos is reduced. This mass range has, however,
already been excluded by accelerator experiments. For very high masses the
number density of neutralinos, and therefore the event rate, becomes small.
This is not a problem as problematically large masses are excluded by
theoretical arguments as previously discussed.

\begin{figure}[h]
\centering
\epsfxsize=3.8in\hspace{0in}\epsffile{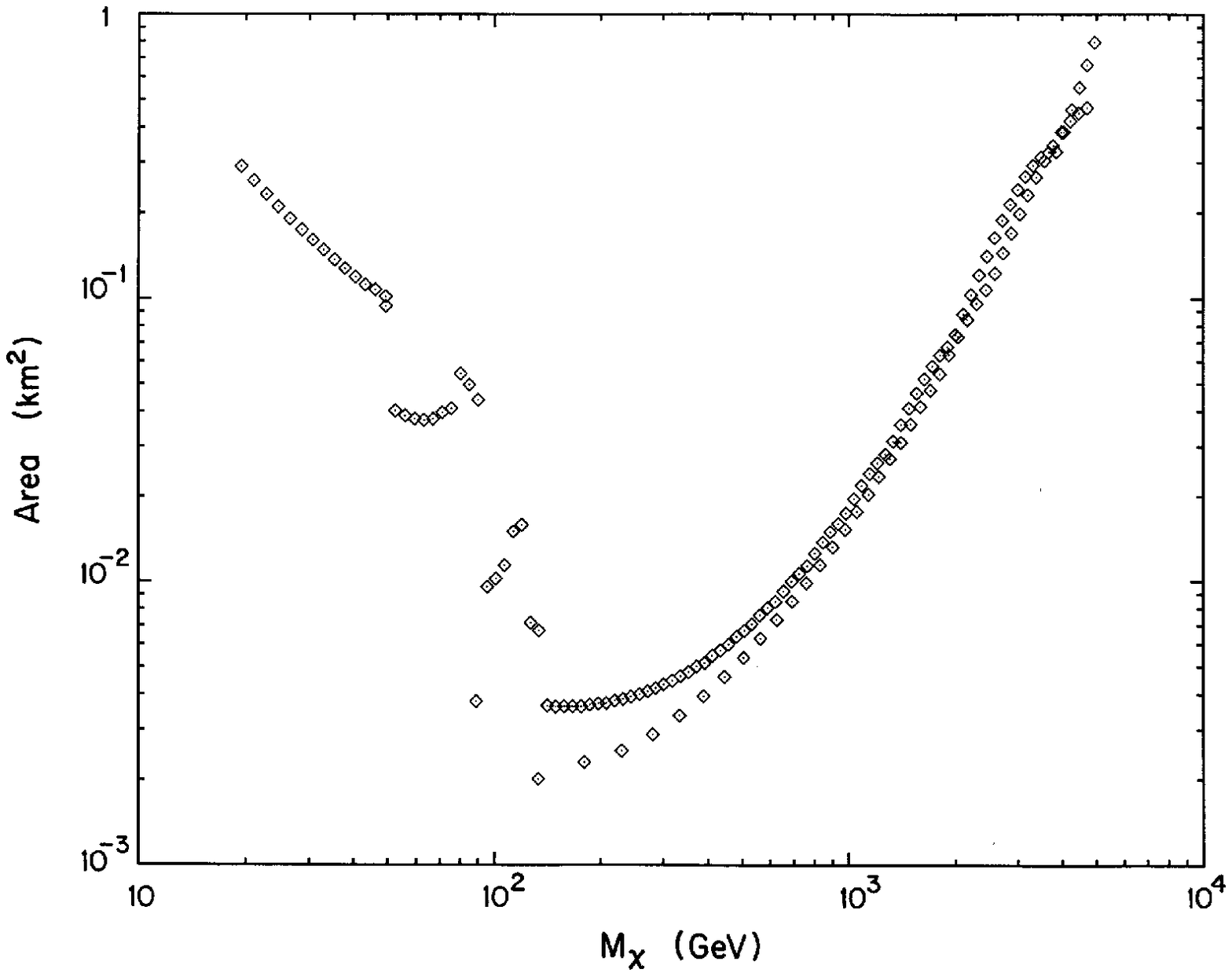}

\parbox{5.5in}{\footnotesize Fig.~5. As a function of the neutrino mass we show
the telescope size required to be sensitive at the one event per year level. We
fix $\tan\beta=2,\ \alpha_{H_2}=0.1$. The two branches correspond to the two
solutions for fixed $\alpha_{H_2}$.}
\end{figure}

The same results are shown in Fig.~6(a) as contours in the $M, \mu$ plane which
denote the neutrino detection area required for observation of 1~event per
year. Clearly the $10^5\,\rm m^2$ contour covers the parameter space. The
problematic large $\mu, M_2$-region does not really represent a problem as its
parameters lead to values of the matter density $\Omega$ exceeding unity as
shown in the accompanying Fig.~6(b).

\begin{figure}[t]
\centering
\epsfxsize=3.175in\hspace{0in}\epsffile{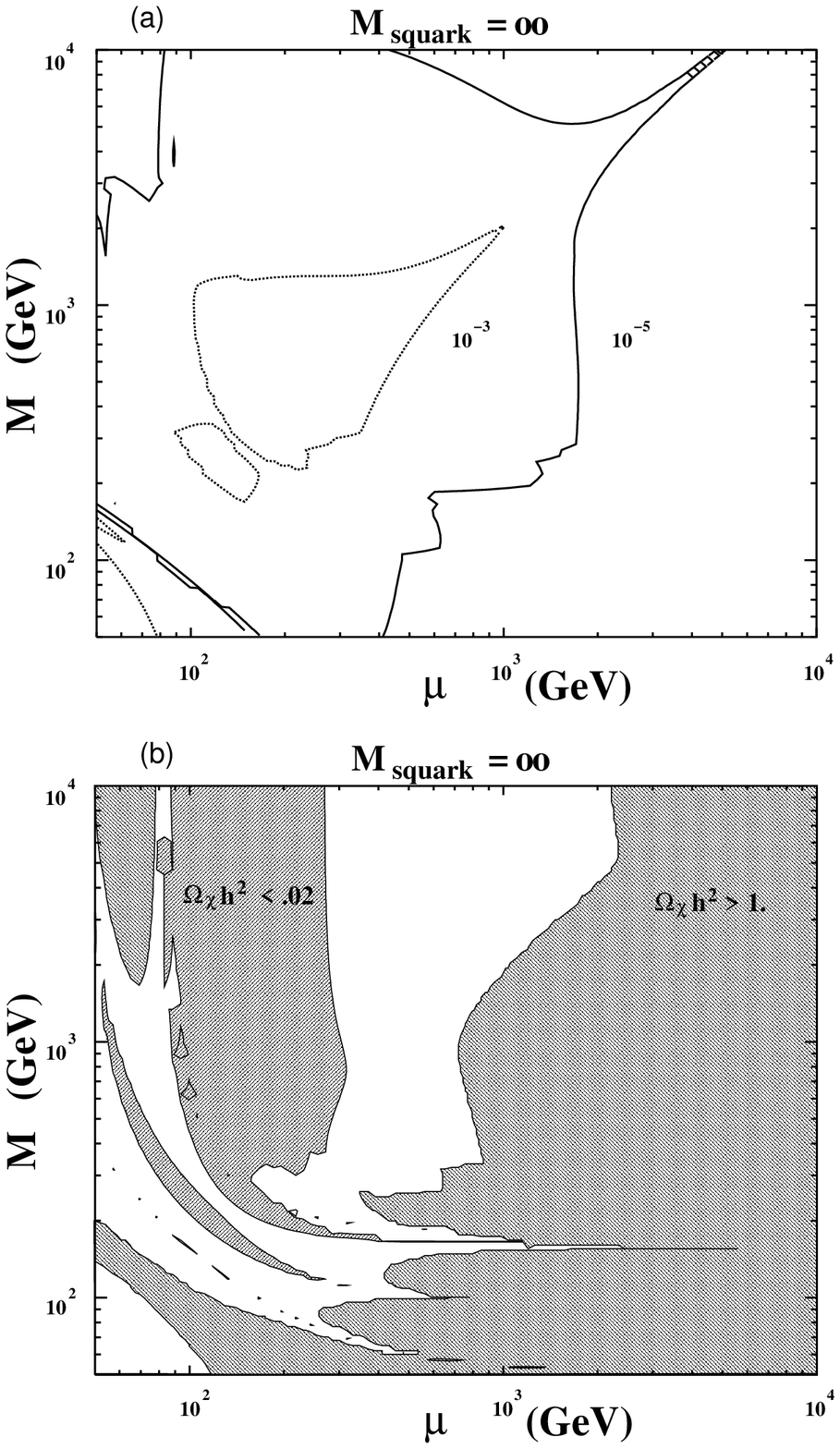}

\medskip
\parbox{5.5in}{\footnotesize Fig.~6. In the $M_2,\mu$ plane for $M_{\tilde
q}=\infty$, (a)~contours
of constant detection rate  (events m$^{-2}$ yr$^{-1}$) and (b)~regions of
$\Omega_\chi h^2 > 1$ and $\Omega_\chi h^2 < 0.02$ which are ruled out by
cosmological considerations.}
\end{figure}

A realistic evaluation of the reach of an underground detector
requires more than counting events per year. Realistic simulations of
statistics and systematics must be done. Also a more complete mapping of the
MSSM parameter space is required. For those interested we refer to
reference~\ref{Halzen}.

We conclude that the natural scale of the complete SUSY detector
is two orders of magnitude larger than the $10^4\,$m$^2$ area of present
detectors, i.e.\ $1\,$km$^2$. This is the size next-generation neutrino
telescopes already under discussion.\cite{Learned} Such an instrument can be
built with roughly the number of phototubes of the SNO experiment in Canada and
the
budget of the Superkamiokande experiment in Japan. We discuss this next.

\section{High Energy Neutrino Telescopes}

High-energy neutrino telescopes are multi-purpose instruments which can make
contributions to astronomy, astrophysics and particle physics. It is
intriguing that all of these missions, including the search for dark matter
discussed here, point to the necessity of building 1~km$^3$
detectors.\cite{Learned} We close with a discussion of the possibility of
building a 1~km scale neutrino detector based on the experience gained in
designing the instruments now under construction, specifically AMANDA, Baikal,
DUMAND and NESTOR which we will briefly review.\cite{Learnedprime} One can
confidently predict that such a telescope can be constructed at a reasonable
cost, e.g.\ a cost similar to Superkamiokande,\cite{Suzuki} to which it is
complimentary in the sense that its volume is over four orders of magnitude
larger while its threshold is in the GeV, rather than the MeV range. The
threshold is in the 2--10~GeV energy range for AMANDA and is about 10~GeV for
DUMAND. Relative to a 1~km scale detector, the experiments under construction
are only ``few" percent prototypes. Yet, using natural water or ice as a
detection medium, these neutrino detectors can be deployed at roughly 1\% of
the cost of conventional accelerator-based neutrino detectors which use
shielding and some variety of tracking chambers.  It is thus not hard to
believe that the Cherenkov detectors can be extended to a larger scale at
reasonable cost.

Detectors presently under construction have a nominal effective area of
$10^4$~m$^2$. Baikal is presently operating 36 optical modules and the South
Pole AMANDA experiment started operating 4 strings with
20 optical modules each in January 94. The first generation telescopes will
consist of roughly 200 optical modules (OM) sensing the Cherenkov light of
cosmic muons. The experimental advantages and challenges are different for
each experiment and, in this sense, they nicely complement one another.
Briefly,

\begin{itemize}
\item  DUMAND will be positioned under 4.5~km of ocean water, below most
biological
activity and well shielded from cosmic ray muon backgrounds. One nuisance of
the ocean is the background light resulting from radioactive decays, mostly
K$^{40}$, plus some bioluminescence, yielding an OM noise rate of
60~kHz. Deep ocean water is, on the other hand very clear with
an attenuation length of order 40~m in the blue. The deep ocean is a
difficult location for access and service, not at all like a laboratory
experiment.  Detection equipment must be built to high reliability standards,
and the data must be transmitted to the shore station for processing.  It has
required years to develop the necessary technology and learn to work in an
environment foreign to high-energy physics experimentation, but hopefully this
will be accomplished satisfactorily.

\item AMANDA is operating in deep clear ice with an attenuation length in
excess of 60~m. Although residual bubbles are found at depth as large as 1~km,
their density decreases rapidly with depth. Ice at the South Pole should be
bubble-free below 1100-1300~m as it is in other polar regions.\cite{Goobar}.
The ice provides a
convenient mechanical support for the detector. The immediate advantage is
that all electronics can be positioned at the surface. Only the optical
modules are deployed into the deep ice. Polar ice is a sterile medium with a
concentration of radioactive elements reduced by more than $10^{-4}$ compared
to sea or lake water. The low background results in an improved sensitivity
which allows for the detection of high energy muons with very simple trigger
schemes which are implemented by off-the-shelf electronics. Being positioned
under only 1~km of ice it is operating in a cosmic ray muon background which
is over 100 times larger than DUMAND. The challenge is to reject the
down-going muon background relative to the up-coming neutrino-induced muons by
a factor larger than $10^6$. The group claims to have met this challenge with
an up/down rejection which is at present superior to that of the deep
detectors. The task is, of course, facilitated by the low
background noise. The polar environment is difficult as well, with
restricted access and one-shot deployment of photomultiplier strings. The
technology has, however, been satisfactorily demonstrated with the deployment
of the first 4 strings. It is now clear that the hot water drilling technique
can be used to deploy OM's larger than the 8~inch photomultiplier tubes now
used to any depth in the 3~km deep ice cover.

\item NESTOR is similar to DUMAND, being placed in the deep ocean (the
Mediterranean), except for two critical differences.  Half of its optical
modules point up, half down. The angular response of the detector is being
tuned to be much more isotropic than either AMANDA or DUMAND, which will give
it advantages in, for instance, the study of neutrino oscillations. Secondly,
NESTOR will have a higher density of photocathode (in some substantial volume)
than the other detectors, and will be able to make local coincidences on lower
energy events, even perhaps down to the supernova energy range (tens of
MeV).

\item BAIKAL shares the shallow depth with AMANDA, and has half its optical
modules pointing up like NESTOR. It is in a lake with 1.4~km
bottom, so it cannot expand downwards and will have to grow horizontally.
Optical backgrounds similar in magnitude to ocean water have been discovered
in Lake Baikal. The Baikal group has been operating for one year an array with
18 Quasar photomultiplier (a Russian-made 15~inch tube) units in April 1993,
and may well count the first neutrinos in a natural water Cherenkov detector.

\item Other detectors have been proposed for near surface lakes or ponds
(e.g.\break GRANDE, LENA, NET, PAN and the Blue Lake Project), but at this time
none are in construction.\cite{Learnedprime}  These detectors all would have
the
great advantage of accessibility and ability for dual use as extensive air
shower detectors, but suffer from the $10^{10}$--10$^{11}$ down-to-up ratio of
muons, and face great civil engineering costs (for water systems and
light-tight containers). Even if any of these are built it would seem that the
costs may be too large to contemplate a full-km scale detector.
\end{itemize}

In summary, there are four major experiments proceeding with construction,
each of which have different strengths and face different challenges. For the
construction of a 1~km scale detector one can imagine any of
the above detectors being the basic building block for the ultimate 1~km$^3$
telescope. The redesigned AMANDA detector (with spacings optimized to the
attenuation length in excess of 60~m), for example, consists of 5 strings on a
60~meter radius circle with a string at the center (referred to as a $1+5$
configuration). Each string contains 13 OMs separated by 15-20~m. Its effective
volume is just below $10^7~m^3$. Imagine
AMANDA ``supermodules'' which are obtained by extending the basic string
length (and module count per string) by a factor close to 4.  Supermodules
would
then consist of $1+5$ strings with, on each string, 51 OMs separated by
20~meters for a length of 1~km. A 1~km scale detector then might
consist of a $1+7+7$ configuration of supermodules, with the 7 supermodules
distributed on a circle of radius 250~m and 7 more on a circle of 500~m. The
full detector contains 4590 phototubes which is less than the 7000 used in the
SNO detector. Such a detector can be operated in a dual mode:

\begin{itemize}
\item it obviously consists of roughly $4\times15$ the presently designed
AMANDA array, leading to an effective volume of $\sim6\times10^8$~km$^3$.
Importantly, the characteristics of the detector,
including low threshold in the GeV-energy range, are the same as those of the
AMANDA array module.

\item the $1+7+7$ supermodule configuration, looked at as a whole, instruments
a 1~km$^3$ cylinder with diameter and height of 1000~m with optical modules.
High-energy
muons will be superbly reconstructed as they can produce triggers in 2 or more
of the modules spaced by large distance.  Reaching more than one supermodule
requires 50~GeV energy to cross 250~m. We note
that this is the
energy for which a neutrino telescope has optimal sensitivity to a typical
$E^{-2}$ source (background falls with threshold energy, and until about
1~TeV little signal is lost).
 \end{itemize}

\begin{figure}[h]
\centering
\epsfxsize=5in\hspace{0in}\epsffile{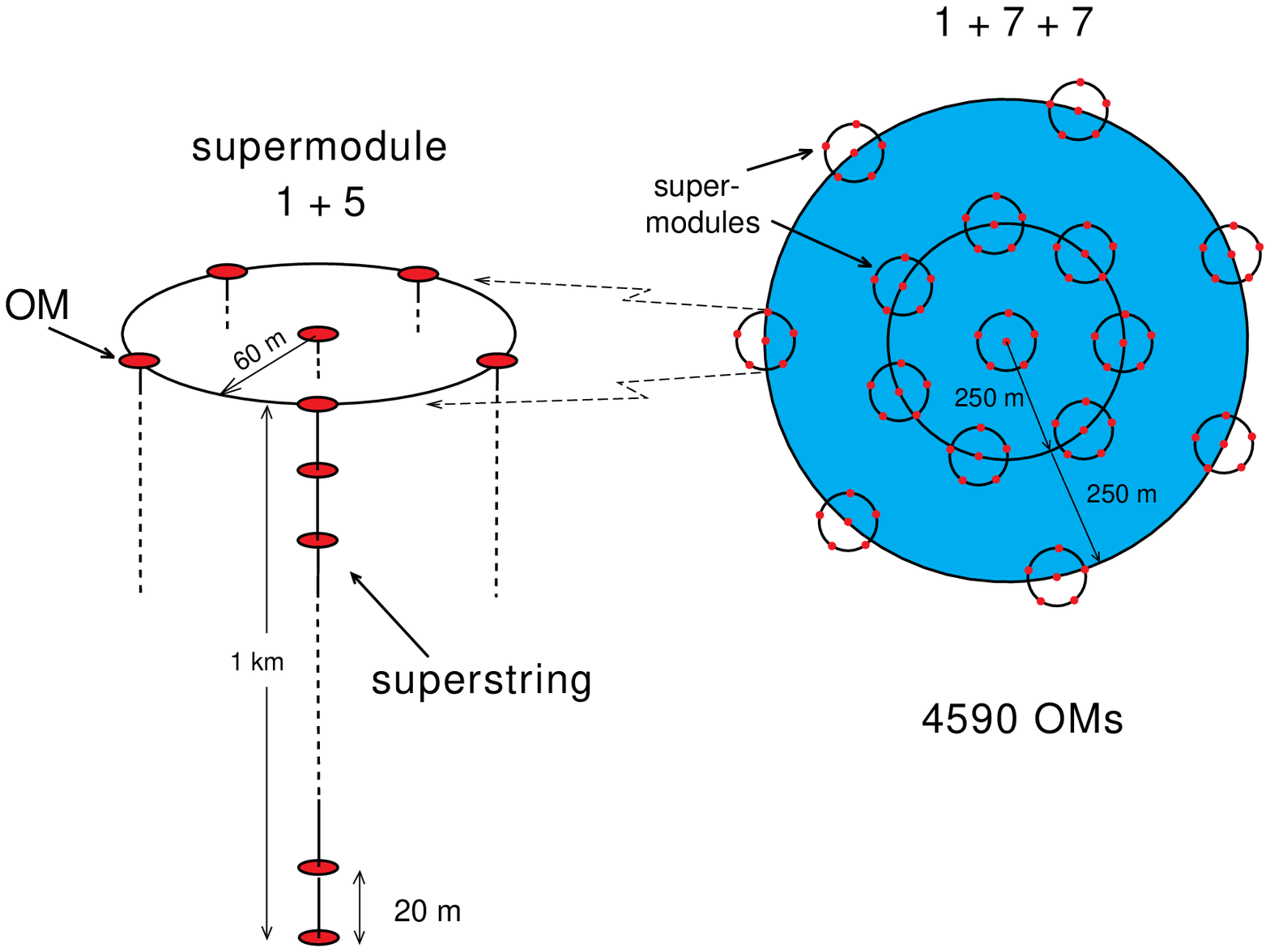}

\footnotesize Fig.~7
\end{figure}

\break
\noindent
Alternate methods to reach the 1~km scale have been discussed by Learned and
Roberts.\cite{Roberts}

How realistic are the construction costs for such a detector?  AMANDA's
strings (with 10 OMs) cost \$150,000 including deployment. By naive scaling the
final cost of
the postulated $1+7+7$ array of supermodules is of order \$50 million and still
below that of Superkamiokande (with $11{,}200 \times 20$~inch photomultiplier
tubes in a 40~m diameter by 40~m high stainless steel tank in a deep mine). It
is clear that the naive estimate makes several approximations over- and
underestimating the actual cost.

At the 1~km$^2$ size it seems inescapable that supersymmetry will be found or,
alternatively, will be forced to ``escape'' in rather special regions of its
vast parameter space. In the latter case the appeal of this beautiful
theoretical idea would be
diminished.

\section*{Acknowledgements}

We would like to thank Steve Barwick, Manuel Drees, Concha Gonzalez-Garcia and
Marc Kamionkowski for useful conversations. This research was supported in
part by the U.S.~Department of Energy under Contracts No.~DE-FG02-91ER40626
and No.~DE-AC02-76ER00881, in part by the Texas Research National Commission
under Grant No.~RGFY9173, and in part by the University of Wisconsin Research
Committee with funds granted by the Wisconsin Alumni Research Foundation.

\section*{References}
\frenchspacing
\begin{enumerate}
\addtolength{\itemsep}{-.05in}

\item\label{Seckel}
J.~R.~Primack, B.~Sadoulet, and D.~Seckel,
 Ann. Rev. Nucl. Part. Sci. {\bf B38} (1988) 751.

\item\label{Berezinsky}
V.~S.~Berezinsky,
 {\it Proc.\ of the Fourth
International Symposium on Neutrino Telescopes}, Venice (1992),
ed.\ by M.~Baldo-Ceolin.

\item\label{Haber}
H.~E.~Haber and G.~L.~Kane,
 Phys. Rep. {\bf 117}, 75 (1985).

\item\label{Drees}
M.~Drees and M.~M.~Nojiri,
 Phys.\ Rev.\ {\bf D47}, 376 (1993).

\item\label{Gaisser}
 For a recent review, see T.~K.~Gaisser {\it et al.}, {\it The Astrophysics and
Particle Physics of High Energy Cosmic Radiation}, National Research Council
Research Briefing, Washington (1994).

\item\label{Kamionkowski}
 For updated calculations and references, see G.~Jungman and M.~Kamionkowski,
Princeton Preprint IASSNS-HEP-93/54, 1993.

\item\label{Turner}
 M. Kamionkowski and M. S. Turner, Phys.\ Rev.\ {\bf D43}, 1774 (1991) and
references therein.

\item\label{Halzen}
 F.~Halzen, M.~Kamionkowski, and T.~Stelzer, Phys.\ Rev.\ {\bf
D45}, 4439 (1992).

\item\label{Bottino}
 A.~Bottino {\it et al.}
Mod. Phys. Lett. A {\bf 7}, 733 (1992).

\item\label{Gould}
 A. Gould,
 Astrophys. J. {\bf 321}, 571 (1987);
Astrophys.~J. {\bf 368}, 610 (1991); Astrophys.~J. in
press (1991).

\item \label{Learned}
 F.~Halzen and J.G.~Learned, {\it Proc.\ of the Fifth
International Symposium on Neutrino Telescopes}, Venice (1993),
ed.\ by  M.~Baldo-Ceolin.

\item\label{Mori}
 M. Mori {\em et al.}, KEK Preprint 91-62; N. Sato {\em et al.},
Phys. Rev. {\bf D44}, 2220 (1991).

\item\label{LoSecco}
 IMB Collaboration: J. M. LoSecco {\em et al.}, Phys. Lett.  {\bf
B188}, 388 (1987); R. Svoboda {\em et al.}, Astrophys.~J. {\bf 315}, 420
(1987).

\item\label{Learnedprime}
 See recent reports on these experiments in the following reference, and a
critical summary of the various projects in J.~G.~Learned, {\it Proceedings of
the European Cosmic Ray Symposium}, Geneva, Switzerland, July 1992, ed. by
P.~Grieder and B.~Pattison,  Nucl.\ Phys.\ B (1993); see also presentations in
{\it Proceedings of the High Energy Neutrino Astrophysics Workshop}, ed. by
V.~J.~Stenger, J.~G.~Learned, S.~Pakvasa, and X.~Tata, World Scientific,
Singapore (1992).

\item\label{Suzuki}
 Y.~Suzuki, {\it Proceedings of the 3$^{rd}$ International
Workshop on Neutrino Telescopes}, Venice, March 1992, ed.\ by M.~Baldo-Ceolin,
Venice (1992).

\item\label{Goobar}
AMANDA-collaboration, Nature (submitted)

\item\label{Roberts} J.~G.~Learned and A.~Roberts, {\it Proceedings of the
23$^{rd}$ International Cosmic Ray Conference}, Calgary, Canada (1993).

\end{enumerate}
\end{document}